# Direct imaging of hot spots in $Bi_2Sr_2CaCu_2O_{8+\delta}$ mesa terahertz sources

T. M. Benseman,[1] A. E. Koshelev,[1] W.-K. Kwok,[1] U. Welp,[1,a] V. K. Vlasko-Vlasov,[1] K. Kadowaki,[2] H. Minami,[2] C. Watanabe[2]

[1]*Materials Science Division, Argonne National Laboratory, Argonne, Illinois, 60439, USA*

[2]*Institute for Materials Science, University of Tsukuba, Ibaraki 305-8753, Japan*

Stacks of intrinsic Josephson junctions (IJJs) made from high-temperature superconductors such as $Bi_2Sr_2CaCu_2O_{8+\delta}$ (Bi-2212) are a promising source of coherent continuous-wave terahertz radiation. It is thought that at electrical bias conditions under which THz-emission occurs hot spots may form due to resistive self-heating, and that these spots may be highly beneficial for the generation of high levels of THz power. Here we perform an imaging study of the temperature distribution at the surface of BSCCO stacks utilizing the temperature-dependent 612nm fluorescence line of $Eu^{3+}$ in a europium chelate. The images directly reveal a highly non-uniform temperature distribution in which the temperature in the middle of the stack can exceed the superconducting transition temperature by tens of Kelvin under biasing conditions typical for THz-emission.

---

[a] Author to whom correspondence should be addressed. Electronic mail: welp@anl.gov

# I. INTRODUCTION

Mesas fabricated from extremely anisotropic high-temperature superconductors such as $Bi_2Sr_2CaCu_2O_{8+\delta}$ (Bi-2212) act as stacks of intrinsic Josephson junctions (IJJs) [1]. For certain mesa geometries, such devices can be biased to generate significant far-field terahertz radiation power, when the Josephson frequency $f_J = 2eV_{junction}/h$ matches the cavity resonance frequency $f_{cavity} = c/2nw$; here, $w$ is the width of the mesa, $n$ the $c$-axis refractive index of the material, $c$ the vacuum speed of light and $V_{junction}$ the voltage per junction [2]. These results show that Bi-2212 mesas are a promising source of coherent radiation in the 'terahertz gap' frequency range from 0.5 to 1.3 THz, in which few other bright, compact sources are available [3].

Extensive studies have shown that due to the low thermal conductivity along the crystallographic $c$-axis, Bi-2212 mesas are prone to self-heating when electrically biased to current-voltage conditions at which THz-emission is expected [4-6]. This problem is particularly serious in the relatively large mesas commonly used for THz-generation. Low-temperature scanning laser microscopy (LTSLM) experiments in which a modulated laser pulse induces a local variation in the $c$-axis resistivity and/or critical current of Bi-2212 indicate that the self-heating is not uniform, but rather takes the form of one or more localized hot spots [7-9]. This is broadly consistent with recent thermal images of the surface temperature of BSCCO mesas [10] and with theoretical calculations for Bi-2212 mesas of this type, which predict non-uniform self-heating of the order of tens of Kelvin [5, 6]. Furthermore, it is possible that hot spots form a shunt between the stacked Josephson junctions thereby promoting their synchronization [5, 11]. This is consistent with studies, which find a narrower THz radiation line width from mesas with comparatively high DC power dissipation [12]. Furthermore, it was observed that mesas covered with thick electrodes that homogenize the temperature do not emit THz-radiation [13]. However, recent thermal images [10] of the surface temperature indicate that there is no correlation between the THz-emission and the hot spot. Understanding the interaction between non-uniform temperature and electromagnetic modes may therefore be essential for the design of Bi-2212 mesas capable of generating technologically interesting levels of THz power.

LTSLM is sensitive to mesa temperature via the strong T-dependence of the *c*-axis resistivity, $\partial \rho_c(x, y)/\partial T(x, y)$, in under- and optimally-doped Bi-2212 mesas, when at least part of the area of the junctions is below $T_c$, but switched into the resistive state. However, it is also sensitive to local variation in the electric field in the mesa, and is thus neither a pure nor a direct measurement of mesa self-heating. It is thus helpful to have a direct microscopic probe of mesa temperature, not only for the study of hot spots, but also to aid interpretation of features seen by LTSLM, such as standing wave patterns running the length of the mesa [7, 8].

## II. EXPERIMENTAL

Here we describe the application of a fast, simple, and economical technique for directly imaging surface temperature of a mesa, first employed by Kolodner *et al.* [14]. It is based on the strong temperature dependence of the optical luminescence of the Eu-coordination complex europium thenoyltrifluoroacetonate (EuTFC). The organic ligands in this complex effectively absorb UV light in a broad band around 345 nm. The energy is transferred radiation-less via intra-molecular excitations to the $Eu^{3+}$ ion, which returns the complex to its ground state through the emission of a luminescence photon at 612 nm. The strong temperature dependence arises from the energy transfer process [15] making for a sensitive thermal probe of an object coated with this material. Previously, variations of this technique employing small concentrations of Eu-chelates dissolved in polymer films have been used to image the onset of dissipation in superconducting microwave filters [16] and in micro-bridges [17]. In this work we use films of EuTFC directly sublimated onto the surface of Bi-2212 mesas, which significantly improves the sensitivity and reduces the image acquisition time.

The Bi-2212 mesa samples investigated here were first characterized for THz emission (see [18] for a description of the optical setup used) and then coated with approximately 200nm of EuTFC. The sample is mounted in a helium flow cryostat with an optical window suitable for microscopy. To excite the fluorescence, the sample is illuminated at an angle of approximately 20 degrees incidence to its surface (to minimize direct reflection of incident light) using a mercury arc lamp (X-cite 120) fitted with a 480nm short-pass filter. The fluorescent signal from the EuTFC coating is imaged using an optical microscope with a cooled CCD camera and an optical filter with a narrow pass-band, centered at 612nm. Using this setup, a fluorescent image of the sample can be obtained, with ~1 micron spatial resolution, and temporal resolution limited by the exposure time of 200 ms. From the CCD count ratio of an image collected with power applied to the mesa to one collected with the sample in thermal equilibrium, a fluorescent image may be obtained which is corrected to first order for any sample-dependent variations in background reflectivity. Such images may be directly converted to sample temperature by varying the bath temperature with zero power applied to the device, thus obtaining a temperature calibration curve for the fluorescence intensity. Since the fluorescence efficiency and the intensity of directly reflected light both depend on the sample surface, a calibration curve must be determined for each surface material, namely Au and Bi-2212 in the case of this experiment as shown in Fig. 2. The fluorescence intensity is close to linear in T over the range of interest, giving a sensitivity of approximately 4.5 K·μm·s$^{1/2}$, i.e. around 1K for the 200 ms collection time and integration area of 10 × 10 μm employed here.

## III. RESULTS and DSICUSSION

Fig. 3a shows the temperature dependence of the c-axis resistance of a BSCCO-mesa of dimensions 300 × 60 × 0.8 microns, containing approximately 530 junctions, and covered with a 100nm gold top electrode. Current is supplied by a lithographically patterned gold current track, 500nm thick. The rise in $R_c$ on reducing the temperature towards $T_c$ and the sharp resistance drop at $T_c \sim 87$ K are typical for optimally doped BSCCO. The inset of Fig. 3a shows current-voltage characteristic measured on decreasing current bias at temperature intervals of 5 K between 10 K and 100 K. The pronounced back-bending of the IV-curve, particularly at the lowest temperatures, is commonly seen on this type of sample [4-6] and is the hallmark of strong self-heating. At temperatures below 45 K re-trapping occurs at currents low enough such that the c-axis resistance can be estimated from the sub-gap voltage as indicated by the solid lines. The results, included in the main panel of Fig. 3a as blue squares, reveal that at low temperatures the c-axis resistance saturates and increase approximately linearly with decreasing temperature [19] at a rate of $\sim 7.4$ $\Omega$/K rather than exponentially [20]. Fig. 3b shows the current-voltage characteristic at 61 K where the strongest THz-emission was observed for this sample together with the emission power as function of mesa voltage. Bias currents for which thermal images were acquired are indicated.

Figure 4a and 4b respectively show a conventional optical micrograph of the mesa and a thermal fluorescent image taken under the conditions at which the mesa's THz output power is maximized, namely 61K and 20mA. We observe a strongly non-uniform temperature distribution at the surface of the mesa, reaching nearly 20K above the bath temperature at the center of the mesa. The evolution of the temperature distribution and of the temperature profiles (taken along the dashed lines indicated in Fig. 4a) upon increasing bias current is shown in Fig. 5. A broad temperature maximum appears in the center of the mesa, in good qualitative agreement with numerical solutions of the heat diffusion equations [5, 6]. The sharp, step-like feature visible in the middle of the longitudinal profiles (Fig. 5a) marks the edge of the Au-current lead sputtered onto the mesa. This temperature drop by more than 10 K (at high bias) indicates that thick Au-contacts on the top of the mesa are efficient in removing dissipated heat. The narrow 'hot' lines observed at the edges of the mesa (clearly seen in the transverse profiles shown in Fig. 5b) and of the Au current track are artifacts due to scattering of incident light at these edges. For similar reasons, the outline of the other (not energized) mesa can also be seen in the thermal images. At bias currents of 60 mA and above, the entire mesa self-heats above $T_c$, even though the current-voltage characteristics is still highly non-linear. Although suggestive, it is at present not clear whether the oscillatory temperature profiles seen at the highest bias currents are a signature of the standing wave patterns reported previously in LTSLM images [7, 8].

The S-shaped current-voltage characteristics – brought about by the strong temperature dependence of the c-axis resistivity – is a consequence of the strong self-heating seen in Figs. 3 and 4. Quite generally, a material displaying sections of negative differential resistance in their IV-curves is susceptible to electric instability [21]. In our case, the S-shaped or current-controlled IV-curve allows for multi-valued currents, which could lead to the formation of domains in the material characterized by different values of the current density. This phenomenon of current-filamentation has been investigated in the context of the electro-thermal breakdown of semiconductors [22, 23] where the current is concentrated in regions of the sample that are significantly smaller than the sample size. Here, the very steep (i.e., exponential) decrease of the resistance of the semiconductor with temperature causes positive feedback resulting in an instability and the formation of a current filament (domain) that is accompanied by a strongly non-uniform temperature distribution centered on this current filament. The $c$-axis resistivity of our BSCCO mesas increases rapidly with decreasing temperature in the temperature range around $T_c$ (see Fig. 3a). However, the approximately linear increase seen here is much weaker than the exponential rise of the resistance of semiconductors and, therefore, it is currently not clear whether a transition into a domain state of tightly concentrated current and temperature, respectively, can arise. The thermal images obtained here indicate broad, smooth temperature variations. It should be noted though that the Au-contact on top of the current-generation mesas may act as a thermal short, significantly reducing the thermal gradient which might otherwise form across the surface of the mesa.

**IV. CONCLUSION**

In summary, we have demonstrated that thermal fluorescence microscopy can be used to image the non-uniform temperature distribution caused by self-heating in Bi-2212 mesas designed for THz generation. In the temperature range of 60 K to 180 K we achieve a temperature resolution of ~ 1K at a spatial resolution of ~ 10 µm. The temperature profiles obtained here are in qualitative agreement with expectations based on solutions of the heat diffusion equation. The oscillatory temperature distribution observed at high current bias may signal the transition into a domain state; however, samples with thinner Au-contacts are required to further investigate this effect.

**ACKNOWLEDGEMENT**

This research was funded by the Department of Energy, Office of Basic Energy Sciences, under Contract No. DE-AC02-06CH11357, which also funds Argonne's Center for Nanoscale Materials (CNM) where the patterning of the BSCCO mesas was performed.

**References**

[1] R. Kleiner, F. Steinmeyer, G. Kunkel, P. Müller, Phys. Rev. Lett. **68**, 2394 (1992).

[2] L. Ozyuzer, A. E. Koshelev, C. Kurter, N. Gopalsami, Q. Li, M. Tachiki, K. Kadowaki, T. Yamamoto, H. Minami, H. Yamaguchi, T. Tachiki, K. E. Gray, W.-K. Kwok, U. Welp, Science 318, 1291 (2007).

[3] For recent reviews: special issue "T-ray imaging, sensing, & detection" Proceedings of the IEEE **95**, nr. 8 (2007); M. Tonouchi, Nature Photonics **1**, 97 (2007); A. Redo-Sanchez, X.-C. Zhang, IEEE J. Sel. Top. Quantum Electron. **14**, 260 (2008); S. Kumar, IEEE J. Sel. Top. Quantum Electron. **17**, 38 (2011).

[4] C, Kurter, K. E. Gray, J. F. Zasadzinski, L. Ozyuzer, A. E. Koshelev, Q. Li, T. Yamamoto, K. Kadowaki, W.-K. Kwok, M. Tachiki, U. Welp, IEEE Appl. Superconductivity **19**, 428 (2009); C, Kurter, L. Ozyuzer, T. Proslier, J. F. Zasadzinski, D. G. Hinks, K. E. Gray, Phys. Rev. B **81**, 224518 (2010); M. Suzuki, T. Watanabe, A. Matsuda, Phys. Rev. Lett. **82**, 5361 (1999); H. B. Wang, T. Hatano, T. Yamashita, P. H. Wu, P. Muller, Appl. Phys. Lett. **86**, 023504 (2005); J. C. Fenton, C. E. Gough, J. Appl. Phys. **94**, 4665 (2003); V. N. Zavaritsky, Phys. Rev. Lett. **92**, 259701 (2004); V. M. Krasnov, M. Sandberg, I. Zogaj, Phys. Rev. Lett., **94**, 077003, (2005).

[5] A. Yurgens, Phys. Rev. B **83**, 184501 (2011).

[6] B. Gross, S. Guenon, J. Yuan, M. Y. Li, J. Li, A. Ishii, R. G. Mints, T. Hatano, P. H. Wu, D. Koelle, H. B. Wang, R. Kleiner, Phys. Rev. B **86**, 094524 (2012).

[7] H. B. Wang, S. Guenon, J. Yuan, A. Iishi, S. Arisawa, T. Hatano, T. Yamashita, D. Koelle, R. Kleiner, Phys. Rev. Lett. **102**, 017006 (2009).

[8] H. B. Wang, S. Guenon, B. Gross, J. Yuan, Z. G. Jiang, Y. Y. Zhong, M. Gruenzweig, A. Iishi, P. H. Wu, T. Hatano, D. Koelle, R. Kleiner, Phys. Rev. Lett. **105**, 057002 (2010).

[9] S. Guenon, M. Gruenzweig, B. Gross, J. Yuan, Z. G. Jiang, Y. Y. Zhong, M. Y. Li, A. Iishi, P. H. Wu, T. Hatano, R. G. Mints, E. Goldobin, D. Koelle, H. B. Wang, R. Kleiner, Phys. Rev. B **82**, 214506 (2010).

[10] H. Minami, C. Watanabe, K. Sato, S. Sekimoto, T. Yamamoto, T. Kashiwaki, R. A. Klemm, K. Kadowaki, to be published.

[11] M. Li, J. Yuan, N. Kinev, J. Li, B. Gross, S. Guenon, A. Ishii, K. Hirata, T. Hatano, D. Koelle, R. Kleiner, V. P. Koshelets, H. B. Wang, P. H. Wu, Phys. Rev. B **86**, 060505 (2012).

[12] A. Larkin, A. Varlamov "Theory of Fluctuations in Superconductors" (Clarendon Press, Oxford, 2005), p. 304.

[13] I. Kakeya, Y. Omukai, T. Yamamoto, K. Kadowaki, M. Suzuki, Appl. Phys. Lett. **100**, 242603 (2012).

[14] P. Kolodner, J. A. Tyson, Appl. Phys. Lett., **40**, 782 (1982); P. Kolodner, J. A. Tyson, Appl. Phys. Lett. **42**, 117 (1983).

[15] G. B. Hadjichristov, S. S. Stanimirov, I. L. Stefanov, I. K. Petkov, Spectrochimica Acta A 69, 443 (2008); S. W. Allison, G. T. Gillies, Rev. Scientific Instr. **68**, 2615 (1997); G. E. Khalil, K. Lau, G. D. Phelan, B. Carlson, M. Gouterman, J. B. Callis, L. R. Dalton, Rev. Sci. Instrum. **75**, 192 (2004); K. Binnemans, Chem. Rev. **109**, 4283 (2009).

[16] G. Hampel, P. Kolodner, P. L. Gammel, P. A. Polakos, E. de Obaldia, P. M. Mankiewich, A. Anderson, R. Slattery, D. Zhang, G. C. Liang, C. F. Shih, Appl. Phys. Lett. **69**, 571 (1996).

[17] O. Haugen, T. H. Johansen, H. Chen, V. Yurchenko, P. Vase, D. Winkler, B. A. Davidson, G. Testa, E. Sarnelli, E. Altshuler, IEEE Trans. Appl. Supercond. **17**, 3215 (2007); S. Niratisairak, O. Haugen, T. H. Johansen, T. Ishibashi, Physica C **468**, 442 (2008).

[18] T. M. Benseman, A. E. Koshelev, K. E. Gray, W.-K. Kwok, U. Welp, K. Kadowaki, M. Tachiki, T. Yamamoto, Phys. Rev. B **84**, 064523 (2011).

[19] J. Takeya, S. Akita, J. Shimoyama, K. Kishio, Physica C **261**, 21 (1996); A. Yurgens, D. Winkler, N. V. Zavaritsky, T. Claeson, Phys. Rev. Lett. **79**, 5122 (1997); Yu. I. Latyshev, T. Yamashita, L. N. Bulaevskii, M. J. Graf, A. V. Balatsky, M. P. Maley, Phys. Rev. Lett. **82**, 5345 (1999).

[20] H. G. Luo, Y. H. Su, T. Xiang, Phys. Rev. B **77**, 014529 (2008).

[21] B. K. Ridley, Proc. Phys. Soc. **82**, 954 (1963); A. V. Gurevich, R. G. Mints, Rev. Mod. Phys. **59**, 941 (1987).

[22] H. Lueder, E. Spenke, Physikalische Zeitschrift **36**, 767 (1936); E. Spenke, Veröffentlichungen aus den Siemens-Werken **15**, 92 (1936).

[23] K. Tsendin, Phys. Status Solidi B **246**, 1831 (2009); N. A. Bogoslovskiy, K. D. Tsendin, Semiconductors **46**, 559 (2012).

**Figure captions:**

Fig. 1: Diagram of microscope, illumination, and cryostat arrangement employed in this experiment.

Fig. 2: Calibration curves of the temperature dependence of the fluorescence intensity over a Au and BSCCO surface measured with respect to a base temperature of 61 K.

Fig. 3: (a) Temperature dependence of the c-axis resistance, $R_c$, of the BSCCO mesa. The blue symbols are estimates of the c-axis resistance obtained from the sub-gap resistance as deduced from the IV-curves (inset of Fig. 3a). The blue dashed line is an interpolation between these data and the resistance data above $T_c$. (b) Current-voltage characteristic for a mesa at 61 K (blue curve, bottom axis) plotted together with the current dependence of the THz output power (red curve, top axis). Arrows indicate current values for which thermal images are shown in Figure 5. The inset shows a schematic of the BSCCO mesa fabricated onto a BSCCO base crystal using photolithography and Ar-ion milling. The wavy red lines symbolize the THz-radiation emitted from the sides of the mesa.

Fig. 4 (a) Optical micrograph of the THz source mesa used in this experiment. An identical adjacent mesa (not energized in this experiment) is also shown near the top of the image. The Au contacts continue to the right; they are however not visible in this image since they follow an inclined sample surface (see inset in Fig. 3b). (b) Thermal fluorescent image of same area at 61K and 20mA, under which conditions the radiated THz power is maximized (at approximately 58 μW). The difference between the mesas with and without applied power is clearly visible.

Fig. 5 (a) & (b) Temperature profiles measured along the dotted lines in Fig. 4 (a) at a bath temperature of 61K, and increasing bias current. (c) Thermal fluorescent image of the mesa at increasing bias current. Images in (c) are calibrated for Au surfaces, but the temperature profiles in (a) & (b) are obtained using a position-dependent calibration function. Note the difference in color-scale in Fig. 5c and Fig. 4b.

PC    Cooled camera

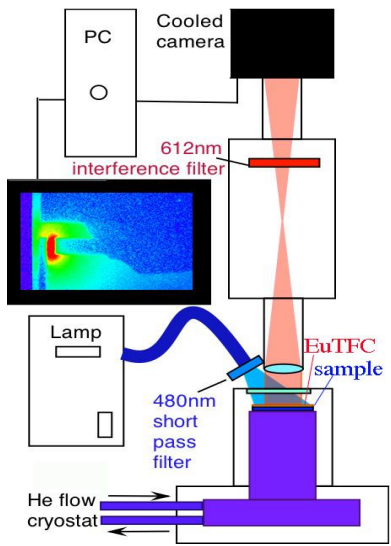

612nm interference filter

Lamp

EuTFC sample

480nm short pass filter

He flow cryostat

Fig. 1

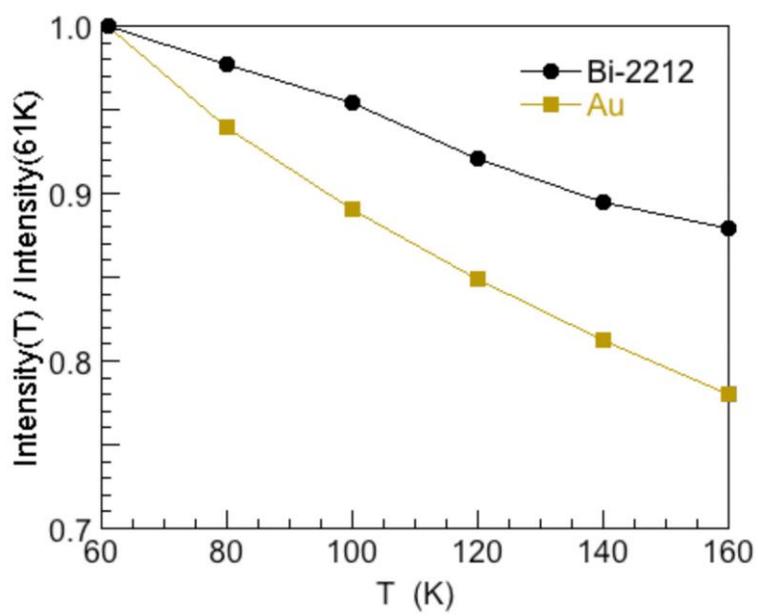

Fig. 2

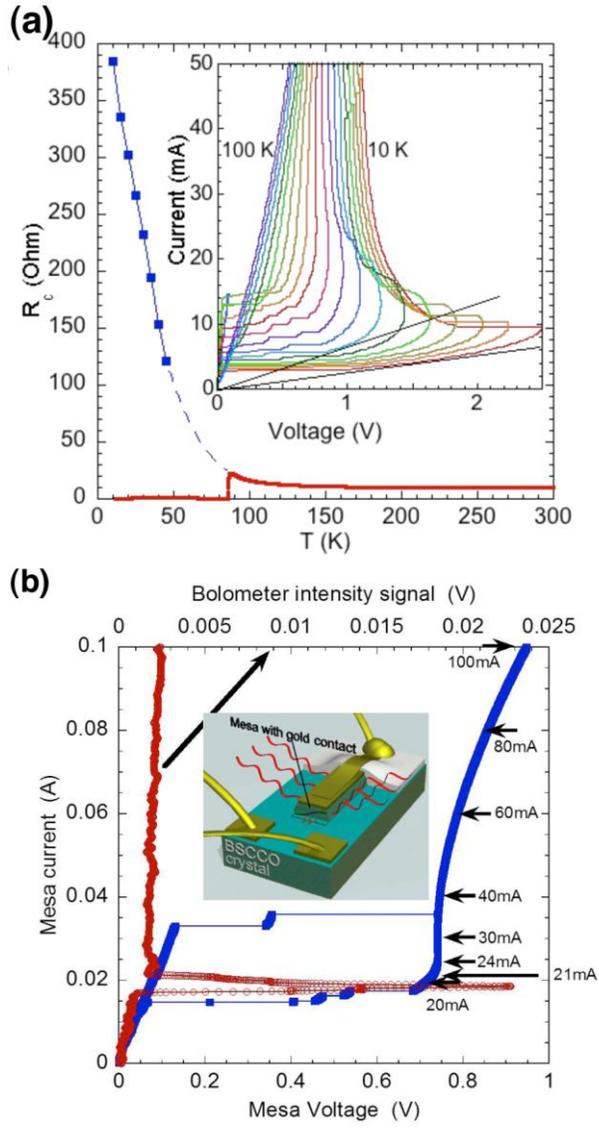

Fig. 3

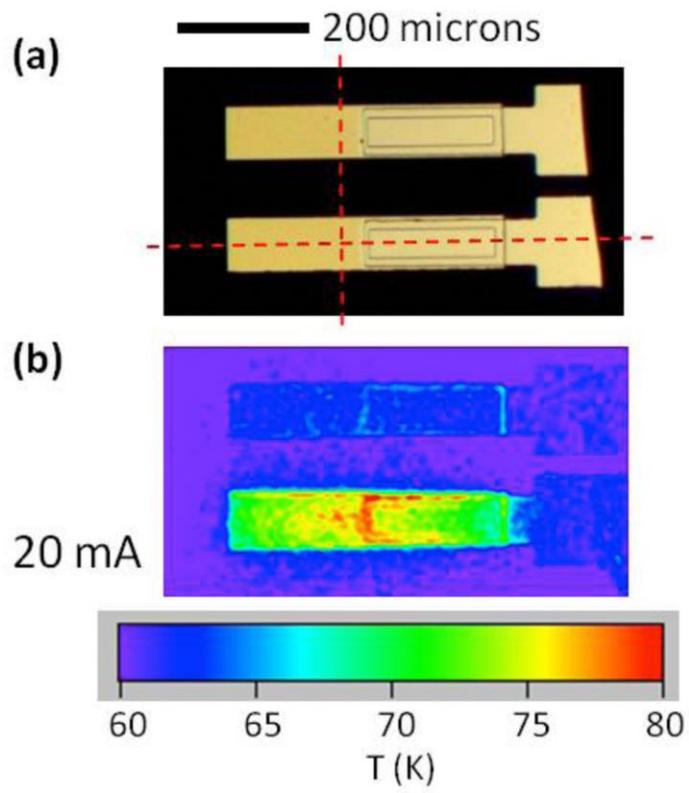

**(a)**

200 microns

**(b)**

20 mA

60   65   70   75   80
T (K)

Fig. 4

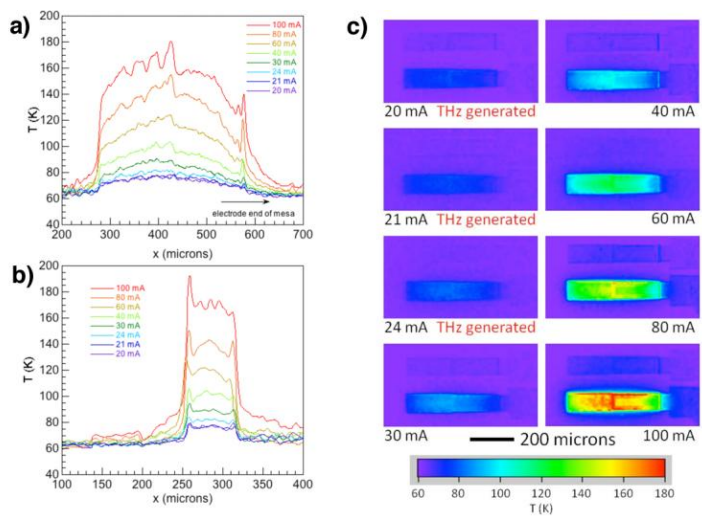

Fig. 5